\documentclass[a4paper]{article}

\newcommand{\be}{\begin{equation}}
\newcommand{\ee}{\end{equation}}
\newcommand{\beq}{\begin{eqnarray}}
\newcommand{\eeq}{\end{eqnarray}}
\newcommand{\no}{\nonumber}
\newcommand{\bea}{\begin{array}}
\newcommand{\eea}{\end{array}}
\newcommand{\lb}{\label}
\newcommand{\mcal}{\mathcal}
\newcommand{\ve}{\varepsilon}

\newcommand{\pp}{\partial}
\newcommand{\im}{\imath}
\newcommand{\ppr}{^{\prime}}

\newcommand{\dtot}{\mbox{d}}
\newcommand{\wtilde}{\widetilde}
\newcommand{\ovv}{\overline}
\newcommand{\Tr}{\raisebox{-4pt}{$\mbox{Tr}\atop {\scriptstyle s,s^{\prime},a,b}$}} 
\newcommand{\trs}{\raisebox{-4pt}{$\mbox{tr}\atop {\scriptstyle s,s^{\prime}}$}} 
\newcommand{\trab}{\raisebox{-4pt}{$\mbox{tr}\atop {\scriptstyle a,b}$}} 

\newcommand{\bvarphi}{\overline{\varphi}}
\newcommand{\balpha}{\overline{\alpha}}

\usepackage[]{graphicx}
\begin{document}
{\Large\bf
\begin{center}
Nonlinear Sigma Model for a Condensate \\ 
Composed of Fermionic Atoms
\end{center}
}
{\small\sf\bf
\begin{center}
Bernhard Mieck \\ 
Department of Physics in Essen, University Duisburg-Essen, \\
Universit\"atsstrasse 5, 45117 Essen, Germany 
  \footnote{B. Mieck\quad E-mail:~\textsf{mieck@theo-phys.uni-essen.de}}
\end{center}
}

\begin{abstract}
A nonlinear sigma model is derived for the time development of a Bose-Einstein condensate
composed of fermionic atoms. Spontaneous symmetry breaking of a $Sp(2)$ symmetry in a
coherent state path integral with anticommuting fields yields Goldstone bosons in a \(Sp(2)\backslash U(2)\) coset space. After a Hubbard-Stratonovich transformation from
the anticommuting fields to a local self-energy matrix with anomalous terms, the assumed short-ranged attractive interaction reduces this symmetry to a \(SO(4)\backslash U(2)\) coset space with only one complex Goldstone field for the singlett pairs of fermions.
This bosonic field for the anomalous term of fermions is separated in a gradient expansion from the density terms. The $U(2)$ invariant density terms are considered as a background field or unchanged interacting Fermi sea in the spontaneous symmetry breaking of the
$SO(4)$ invariant action and appear as coefficients of correlation functions in the nonlinear sigma model for the Goldstone boson. The time development of the condensate composed of fermionic atoms results in a modified Sine-Gordon equation.\newline

\end{abstract}                
\vspace*{1.0cm}

\section{Introduction} 

Experiments of Bose-Einstein condensation (BEC) with bosonic constituents have been
realized under various conditions. In many cases the
Gross-Pitaevskii (GP) equation with a bosonic field $\psi_{\vec{x}}(t)$ as the order
parameter and wavefunction can be applied
\be \lb{1}
\im\hbar\frac{\pp\psi_{\vec{x}}(t)}{\pp t}=-\frac{\hbar^{2}}{2m}\vec{\nabla}^{2}
\psi_{\vec{x}}(t)+u(\vec{x},t)\;\psi_{\vec{x}}(t)+2\sum_{\vec{x}\ppr}
|\psi_{\vec{x}\ppr}(t)|^{2}\;V_{|\vec{x}\ppr-\vec{x}|}\;\psi_{\vec{x}}(t),
\ee
where $u(\vec{x},t)$ refers to a time dependent external potential, including the trap
potential, and $V_{|\vec{x}\ppr-\vec{x}|}$ is a short-ranged interaction \cite{Pit,Peth}. 
Transfering the GP equation (\ref{1}) to the case with fermionic atoms, 
a coherent field equation with anticommuting numbers can be 
introduced where the classical field \(\psi_{\vec{x}}(t)\) is replaced by a
Grassmann-valued field \(\chi_{\vec{x},s}(t)\) with spin \(s=\uparrow,\downarrow\)
\cite{Klau}
\beq \lb{2}
\im\hbar\frac{\pp\chi_{\vec{x},s}(t)}{\pp t}&=&-\frac{\hbar^{2}}{2m}
\vec{\nabla}^{2}\chi_{\vec{x},s}(t)+u(\vec{x},t)\;\chi_{\vec{x},s}(t) + \\ \no &+& 2
\sum_{\vec{x}\ppr,s\ppr} \chi_{\vec{x}\ppr,s\ppr}^{*}(t)\;V_{|\vec{x}\ppr-\vec{x}|}\;
\chi_{\vec{x}\ppr,s\ppr}(t)\;\chi_{\vec{x},s}(t).
\eeq
As its bosonic counterpart, the Grassmann-valued equation is integrable for a contact
interaction and possesses a set of infinite independent integrals of motion \cite{Cau}. 
This has been demonstrated by the method of Lax-pairs and r-matrix methods. 

In this paper an effective nonlinear sigma model of bosonic fields for a condensate composed of fermionic atoms is derived for the following quantum Hamiltonian of Fermi operators \(\psi_{\vec{x},s}\) corresponding to the classical equation (\ref{2}) with anticommuting
fields \(\chi_{\vec{x},s}(t)\) \footnote{The spatial sum \(\sum_{\vec{x}}\ldots\) is dimensionless and is scaled with the system volume so that \(\sum_{\vec{x}}\ldots\) is equivalent to \(\int_{L^{d}}\dtot^{d}x/L^{d}\ldots\).} 
\beq \lb{3}
\mcal{H}(\psi^{+},\psi,t) &=& \sum_{\vec{x},s;\vec{x}\ppr,s\ppr}
\psi_{\vec{x}\ppr,s\ppr}^{+}\;H_{\vec{x}\ppr,s\ppr;\vec{x},s}(t)\;\psi_{\vec{x},s} \\ \no
&+& \sum_{\vec{x},s;\vec{x}\ppr,s\ppr}\psi_{\vec{x}\ppr,s\ppr}^{+}\;
\psi_{\vec{x},s}^{+}\;V_{|\vec{x}\ppr-\vec{x}|}\;\psi_{\vec{x},s}\;
\psi_{\vec{x}\ppr,s\ppr} \\ \no &+& \sum_{\vec{x}}2\; \Big(\wtilde{j}_{\vec{x}}^{*}(t)\;
\psi_{\vec{x},\uparrow}\;\psi_{\vec{x},\downarrow}+
\psi_{\vec{x},\downarrow}^{+}\;\psi_{\vec{x},\uparrow}^{+}\;\wtilde{j}_{\vec{x}}(t)\Big) \\ \lb{4} H_{\vec{x}\ppr,s\ppr;\vec{x},s}(t) &=& \delta_{s\ppr,s}\;\delta_{\vec{x}\ppr,\vec{x}}
\bigg(\frac{\vec{p}^{2}}{2m}+v(\vec{x},t)\bigg) \\ \lb{5}
v(\vec{x},t) &=& u(\vec{x},t)-\mu_{0}\;\; .
\eeq
The external potential \(u(\vec{x},t)\) is shifted by the transformation
\(\exp\{-\im/\hbar\cdot \mu_{0}\;t\}\;\psi_{\vec{x},s}\) with the chemical potential
\(\mu_{0}\). This shift is performed because the time derivative is considered as a 
perturbation in a gradient expansion so that rapid oscillations of the fields
do not appear. Since we also include a time dependence in the external potential
$u(\vec{x},t)$, the chemical potential $\mu_{0}$ is in general not an equlibrium
value and can be extended with an appropriate adiabatic time dependence $\mu(t)$. However, we
assume that the interacting Fermi sea is not strongly perturbed by the time dependence
of $u(\vec{x},t)$. The interaction \(V_{|\vec{x}\ppr-\vec{x}|}<0\) is attractive and has
to be short-ranged in order to obtain local sigma matrices for the self-energies.
Furthermore, the densities of the interacting Fermi sea are regarded
as given background fields which are considered as given coefficients for
the gradients of the bosonic nonlinear sigma model. The spatial gradient expansion for
the nonlinear sigma model is combined with a kind of Thomas-Fermi approximation\cite{Peth}
where the derivative terms of the kinetic energy are taken into account as a perturbation
\cite{Pruis1}-\cite{Pruis3}. 
In the remainder we investigate and assume a BCS like condensation
phenomenon of the atoms, derived from spontaneous symmetry breaking with the source field
\(\wtilde{j}_{\vec{x}}^{*}(t)\), and exclude
the formation of single bosons from bound pairs of atoms because the attractive
potential is taken sufficiently short-ranged. In the case of a box potential with
depth \(V_{0}<0\) and range \(r_{0}\), this means that 
\(m\;|V_{0}|\;r_{0}^{2}/\hbar^{2}<\pi^{2}/8\) has to be sufficiently small \cite{Ball}.
This case with short-ranged attractive interaction is different from the formation of excitons and biexcitons of the long-ranged Coulomb potential in semiconductors where the
Hamiltonian for the strongly bound electrons in semiconductors can be reduced to purely
bosonic operators \cite{Mosk}. A review of the nonlinear sigma model in superconducting
systems with {\it delta-function} correlated disordered potentials and the replica-trick can be found in Refs. \cite{L1,L2}.

A coherent state path integral \cite{Sch}-\cite{Ka} with Grassmann fields \(\chi_{\vec{x},s}(t_{p})=
\chi_{\vec{y}}(t_{p})\) (\(\vec{y}=\{\vec{x},s\}\))\footnote{In the following
$\vec{y}$-vectors refer to the combined spatial vector $\vec{x}$ and spin variable
\(s=\uparrow,\downarrow\) as abbreviation.}
is used on a nonequilibrium
time contour \(\int_{C}\dtot t_{p}\ldots =\int_{-\infty}^{\infty}\dtot t_{+}\ldots +
\int_{\infty}^{-\infty}\dtot t_{-}\ldots\) to express the time development of the
system with Hamiltonian (\ref{3})
\beq \lb{6}
Z[\mcal{J}] &=& \int\dtot[\chi_{\vec{y}}(t_{p})] \\ \no &&
\exp\bigg\{-\frac{\im}{\hbar}\int_{C}\dtot t_{p}\sum_{\vec{y};\vec{y}\ppr}
\chi_{\vec{y}\ppr}^{*}(t_{p})\underbrace{\Big[\delta_{\vec{y}\ppr;\vec{y}}
\Big(-\im\hbar\frac{\pp}{\pp t_{p}}-\im\ve_{p}\Big)+H_{\vec{y}\ppr;\vec{y}}(t_{p})\Big]}_{
\wtilde{\mcal{H}}_{\vec{x}\ppr,s\ppr;\vec{x},s}(t_{p})}
\chi_{\vec{y}}(t_{p})\bigg\} \\ \no &\times &
\exp\bigg\{-\frac{\im}{\hbar}\int_{C}\dtot t_{p}\sum_{\vec{y};\vec{y}\ppr}
\chi_{\vec{y}\ppr}^{*}(t_{p})\;\chi_{\vec{y}}^{*}(t_{p})\;V_{|\vec{x}\ppr-\vec{x}|}\;
\chi_{\vec{y}}(t_{p})\;\chi_{\vec{y}\ppr}(t_{p})\bigg\} \\ \no &\times &
\exp\bigg\{-\frac{\im}{\hbar}\int_{C}\dtot t_{p}\sum_{\vec{x}}\wtilde{j}_{\vec{x}}^{*}(t_{p})\;
\Big(\chi_{\vec{x},\uparrow}(t_{p})\;\chi_{\vec{x},\downarrow}(t_{p})-
\chi_{\vec{x},\downarrow}(t_{p})\;\chi_{\vec{x},\uparrow}(t_{p})\Big)+\mbox{h.c.}\bigg\}
\\ \no &\times & \exp\bigg\{-\frac{\im}{2\hbar}\int_{C}\dtot t_{p_{1}}^{(1)}\;
\dtot t_{p_{2}}^{(2)}\sum_{\vec{y};\vec{y}\ppr}
\eta_{\vec{y}\ppr}^{*}(t_{p_{2}}^{(2)})\;\mcal{J}_{\vec{y}\ppr;\vec{y}}(t_{p_{2}}^{(2)};
t_{p_{1}}^{(1)})\;\eta_{\vec{y}}(t_{p_{1}}^{(1)})\bigg\} \\ \lb{7}
\chi_{\vec{y}}(t_{p}) &=& \chi_{\vec{x},s}(t_{p}) \\ \lb{8}
\eta_{\vec{y}}(t_{p}) &=& \eta_{\vec{x},s}(t_{p}) =
\left(
\bea{c}
\chi_{\vec{x},s}(t_{p}) \\ \chi_{\vec{x},s}^{*}(t_{p})
\eea
\right)_{.}
\eeq
A source field \(\wtilde{j}_{\vec{x}}^{*}(t_{p})\) is applied to generate bosonic pairs
\(\chi_{\vec{x},\uparrow}(t_{p})\;\chi_{\vec{x},\downarrow}(t_{p})\) out of the
vacuum state for spontaneous symmetry breaking of the $U(1)$ invariant one particle
part \(\wtilde{\mcal{H}}\) and the $U(1)$ invariant interaction \cite{Gold,Nambu}. In order to generate observables, 
the source term \(\mcal{J}_{\vec{y}\ppr;\vec{y}}(t_{p_{2}}^{(2)};t_{p_{1}}^{(1)})\)
has to be introduced where the fields \(\chi_{\vec{x},\uparrow}(t_{p})\) and 
\(\chi_{\vec{x},\downarrow}(t_{p})\) have to be combined to the four component vector
\(\eta_{\vec{x},s}(t_{p})\) (\ref{8}) so that pair condensate terms
\(\chi_{\vec{x}\ppr,s\ppr}(t_{p_{2}}^{(2)})\;\chi_{\vec{x},s}(t_{p_{1}}^{(1)})\) can be
obtained by simple differentiation with respect to the matrix
\(\mcal{J}_{\vec{y}\ppr;\vec{y}}(t_{p_{2}}^{(2)};t_{p_{1}}^{(1)})\) \cite{Wen}.
A nonhermitian infinitesimal part \(-\im\;\ve_{p}=-\im\;(\pm\ve)\), (\(p=\pm\)) on the time contour has to be included for the analytic and convergence properties of Green functions,
derived from the coherent state path integral \(Z[\mcal{J}]\) (\ref{6}).

Apart from the commutation relations the field equation, Hamiltonian and coherent
state path integral are formally similar in terms of Grassmann numbers to the
bosonic case \cite{Neg}. Therefore, variations of the action in (\ref{6}) and other
approximations can be performed, however, compared to the condensation of single
bosonic constituents, only a small fraction of fermions can condense near the Fermi
energy \cite{Pit}. This means in terms of a density matrix formulation that expectation
values of densities \(\langle\chi_{\vec{x}\ppr,s\ppr}(t_{p})\;
\chi_{\vec{x},s}^{*}(t_{p})\rangle\) are considerably larger than the pair
condensate \(\langle\chi_{\vec{x}\ppr,s\ppr}(t_{p})\;
\chi_{\vec{x},s}(t_{p})\rangle\). It is the aim of this paper to extract
the various densities and pair condensate functions from the coherent state path
integral and to derive effective equations for the pair condensate composed of fermionic
atoms, in analogy to the GP-equation for bosons (\ref{1}) or fermions (\ref{2}).
The nonlocal order parameter \(\Phi_{\vec{y}\ppr;\vec{y}}(t_{p})\) has a matrix form
\beq \lb{9}
\Phi_{\vec{y}\ppr;\vec{y}}(t_{p})&=&
\left(
\bea{c}
\chi_{\vec{x}\ppr,s\ppr}(t_{p}) \\
\chi^{*}_{\vec{x}\ppr,s\ppr}(t_{p})
\eea\right)\otimes \left(
\bea{cc}
\chi^{*}_{\vec{x},s}(t_{p}),&\chi_{\vec{x},s}(t_{p})
\eea\right) \\ \no &=&
\left(
\bea{cc}
\langle\chi_{\vec{x}\ppr,s\ppr}(t_{p})\;\chi_{\vec{x},s}^{*}(t_{p})\rangle &
\langle\chi_{\vec{x}\ppr,s\ppr}(t_{p})\;\chi_{\vec{x},s}(t_{p})\rangle \\
\langle\chi_{\vec{x}\ppr,s\ppr}^{*}(t_{p})\;\chi_{\vec{x},s}^{*}(t_{p})\rangle &
-\langle\chi_{\vec{x},s}(t_{p})\;\chi_{\vec{x}\ppr,s\ppr}^{*}(t_{p})\rangle 
\eea
\right)_{,}
\eeq
where a doubling of the spin space has to be considered because of the source field
$\wtilde{j}_{\vec{x}}^{*}(t_{p})$ which causes the spontaneous symmetry breaking.
The order parameter (\ref{9}) is invariant under $U(2)$ transformations in spin space
which does not alter the block structure into densities and pair condesates.
The form of the order parameter also allows a global hyperbolic symmetry which combines
densities and pair condensates. A complete symmetry group of the 
path integral is spontaneously broken by the subgroup $U(2)$ for the invariance of the
densities and the source term. This symmetry breaking leads to a nonlinear sigma model after a gradient expansion for the anomalous terms. The various steps for obtaining
the nonlinear sigma model are briefly listed :
\begin{itemize}
\item coherent state path integral
\item transformation of the quartic interaction to densities, anomalous terms
and the order parameter
\item Hubbard-Stratonovich transformation from the fields to the self-energy
and integration over the remaining bilinear anticommuting fields \cite{St,Neg}
\item the short-ranged attractive interaction reduces the $Sp(2)$ symmetry of the
path integral with a {\it spatially nonlocal} self-energy to a {\it spatially local}
self-energy matrix with $SO(4)$ symmetry
\item separation of the self-energy into densities and non-diagonal terms on a coset space 
according to spontaneous breaking of the orthogonal symmetry \(SO(4)\backslash U(2)\) and determination of the measure
\item separation of the coherent state path integral into block diagonal $U(2)$ invariant density matrices and anomalous terms including a gradient expansion
\end{itemize}

\section{Hubbard-Stratonovich transformation and self-energy}

The quartic interaction with the short-ranged two body potential has to be transformed
to a relation with an order parameter similar to \(\Phi_{\vec{y}\ppr;\vec{y}}(t_{p})\)
(\ref{9}). The antisymmetric fields \(\chi_{\vec{x},s}(t_{p})\),
\(\chi_{\vec{x},s}^{*}(t_{p})\) and \(\chi_{\vec{x}\ppr,s\ppr}(t_{p})\),
\(\chi_{\vec{x}\ppr,s\ppr}^{*}(t_{p})\) can be combined in the following even matrices
\(r_{\vec{y}\ppr;\vec{y}}(t_{p})=r_{\vec{x}\ppr,s\ppr;\vec{x},s}(t_{p})\) and
\(\rho_{\vec{y}\ppr;\vec{y}}(t_{p})=\rho_{\vec{x}\ppr,s\ppr;\vec{x},s}(t_{p})\) where
$r$ and $\rho$ are hermitian and antisymmetric, respectively
\beq \lb{10} 
r_{\vec{y}\ppr;\vec{y}}(t_{p}) &=&\chi_{\vec{y}\ppr}(t_{p})\;
\chi_{\vec{y}}^{*}(t_{p}) \\ \lb{11}
r_{\vec{y}\ppr;\vec{y}}^{*}(t_{p}) &=&\chi_{\vec{y}}(t_{p})\;
\chi_{\vec{y}\ppr}^{*}(t_{p}) = r_{\vec{y};\vec{y}\ppr}(t_{p})\rightarrow
r^{+}(t_{p})=r(t_{p}) \\ \lb{12}
\rho_{\vec{y}\ppr;\vec{y}}(t_{p})&=&\chi_{\vec{y}\ppr}(t_{p})\;
\chi_{\vec{y}}(t_{p})=-\chi_{\vec{y}}(t_{p})\;\chi_{\vec{y}\ppr}(t_{p})=
-\rho_{\vec{y};\vec{y}\ppr}(t_{p})\rightarrow
\rho(t_{p})=-\rho^{T}(t_{p})  \\ \lb{13}
\rho_{\vec{y}\ppr;\vec{y}}^{*}(t_{p})&=&\chi_{\vec{y}}^{*}(t_{p})\;
\chi_{\vec{y}\ppr}^{*}(t_{p})=\rho_{\vec{y};\vec{y}\ppr}^{T*}(t_{p})=
\rho_{\vec{y};\vec{y}\ppr}^{+}(t_{p})\; .
\eeq
In terms of the even matrices $r$ and $\rho$, the quartic interaction can be written
in the form of an order parameter \(R_{\vec{x},s;\vec{x}\ppr,s\ppr}^{ab}(t_{p})\)
as in (\ref{9}) with a doubling of the dimension of spin space where the superscripts
$a,\;b=1,2$ refer to the doubling and $s,\;s\ppr=\uparrow,\downarrow$ to the spins
\beq \lb{14}
\lefteqn{
\sum_{\vec{x},s;\vec{x}\ppr,s\ppr}\chi^{*}_{\vec{x}\ppr,s\ppr}(t_{p})\;
\chi_{\vec{x},s}^{*}(t_{p})\;\chi_{\vec{x},s}(t_{p})\;\chi_{\vec{x}\ppr,s\ppr}(t_{p})\;
V_{|\vec{x}\ppr-\vec{x}|} =}  \\ \no &=&\frac{1}{4}\sum_{\vec{y};\vec{y}\ppr}
\Big(\chi^{*}_{\vec{y}\ppr}(t_{p})\;\chi_{\vec{y}\ppr}(t_{p})-
\chi_{\vec{y}\ppr}(t_{p})\;\chi^{*}_{\vec{y}\ppr}(t_{p})\Big)\;
\Big(\chi^{*}_{\vec{y}}(t_{p})\;\chi_{\vec{y}}(t_{p})-
\chi_{\vec{y}}(t_{p})\;\chi^{*}_{\vec{y}}(t_{p})\Big)\;V_{|\vec{x}\ppr-\vec{x}|}
\\ \no &=&
-\frac{1}{4}\sum_{\vec{x},\vec{x}\ppr}V_{|\vec{x}\ppr-\vec{x}|} \Tr
\left[R_{\vec{x},s;\vec{x}\ppr,s\ppr}^{ab}(t_{p})\left(
\bea{cc}
1_{2} & \\
 & -1_{2}
\eea
\right)
R_{\vec{x}\ppr,s\ppr;\vec{x},s}^{ba}(t_{p})\left(
\bea{cc}
1_{2} & \\
 & -1_{2}
\eea
\right)
\right]
\eeq
\beq \lb{15}
R_{\vec{x},s;\vec{x}\ppr,s\ppr}^{ab}(t_{p})&=&
\left(
\bea{c}
\chi_{\vec{x},s}(t_{p}) \\
\chi^{*}_{\vec{x},s}(t_{p})
\eea\right)\otimes \left(
\bea{cc}
\chi^{*}_{\vec{x}\ppr,s\ppr}(t_{p}),&\chi_{\vec{x}\ppr,s\ppr}(t_{p})
\eea\right) \\ \no &=&
\left(
\bea{cc}
r_{\vec{x},s;\vec{x}\ppr,s\ppr}(t_{p}) & \rho_{\vec{x},s;\vec{x}\ppr,s\ppr}(t_{p}) \\
\rho_{\vec{x},s;\vec{x}\ppr,s\ppr}^{+}(t_{p}) & -r_{\vec{x},s;\vec{x}\ppr,s\ppr}^{T}(t_{p}) 
\eea
\right)_{.}
\eeq
Obviously, the quartic interaction can be expressed with the nonlocal order parameter
\(R_{\vec{x},s;\vec{x}\ppr,s\ppr}^{ab}(t_{p})\) so that a global hyperbolic symmetry results with the diagonal matrix \(\kappa=\mbox{diag}(1,1,-1,-1)\) and the matrix $T$ between densities and pair condensates (\(T^{+}\;\kappa\;T=\kappa\)), apart from a $U(2)$ invariance in spin space
\beq \lb{16}
R &\rightarrow & T\;R\;T^{+} \\ \lb{17}
T &=& \left(
\bea{cc}
\sqrt{1+t^{+}t} & t^{+} \\
t & \sqrt{1+tt^{+}}
\eea\right)\hspace*{0.5cm} t:=t_{ss\ppr}\hspace*{0.5cm}t=t^{T} 
\eeq
\be \lb{18}
T^{+}\;\underbrace{\left(
\bea{cc}
1_{2} & \\
 & -1_{2}
\eea\right)}_{\kappa}\;T=
\underbrace{\left(
\bea{cc}
1_{2} & \\
 & -1_{2}
\eea\right)}_{\kappa}\;\; .
\ee
The $2\times 2$ matrix $t_{ss\ppr}$ in spin space of the \(4\times 4\) matrix  $T_{ss\ppr}^{ab}$ has to be complex symmetric and therefore contains 6 real parameters (see appendix \ref{App1}). The hyperbolic symmetry with matrix $T$ and the $U(2)$
invariance in spin space is equivalent to a symplectic symmetry group $Sp(N/2)$ and not
a unitary group as $U(N/2,N/2)$ because of the number of independent parameters which equals ten in the considered case. The symplectic symmetry becomes obvious after an exchange of the first, second with the third, fourth row of the diagonal matrix $\kappa$ and reordering the matrix $T^{+}$ to its transpose $T^{T}$ in relation (\ref{18}) (see appendix \ref{App1}).
Since the matrix $T$ has dimensions \(4\times 4\) and the number of independent parameters
for \(Sp(N/2)\) is \(\frac{1}{2}N(N+1)\), a \(Sp(2)\) invariance is obtained for relation
(\ref{14}). Using the identity for the Hubbard-Stratonovich transformation with the self-energy
matrix \(\Sigma_{\vec{x},s;\vec{x}\ppr,s\ppr}^{ab}(t_{p})\) consisting of commuting
elements only
\be \lb{19}
\Sigma_{\vec{x},s;\vec{x}\ppr,s\ppr}^{ab}(t_{p})=\left(
\bea{cc}
s_{\vec{x},s;\vec{x}\ppr,s\ppr}(t_{p}) & \sigma_{\vec{x},s;\vec{x}\ppr,s\ppr}(t_{p}) \\
\sigma_{\vec{x},s;\vec{x}\ppr,s\ppr}^{+}(t_{p}) & -s_{\vec{x},s;\vec{x}\ppr,s\ppr}^{T}(t_{p}) \eea\right)\hspace*{0.5cm} s=s^{+}\;\;\sigma=-\sigma^{T}\; ,
\ee
the quartic interaction term with $V_{|\vec{x}\ppr-\vec{x}|}$ can be expressed as a
quadratic term of the self-energy and a bilinear product of anticommuting
fields \(\eta_{\vec{x},s}(t_{p})=(\chi_{\vec{x},s}(t_{p}),\chi_{\vec{x},s}^{*}(t_{p}))^{T}\)
\beq \lb{20}
\lefteqn{\exp\bigg\{-\frac{\im}{\hbar}\int_{C}\dtot t_{p}\sum_{\vec{y};\vec{y}\ppr}
\chi_{\vec{y}\ppr}^{*}(t_{p})\;\chi_{\vec{y}}^{*}(t_{p})\;V_{|\vec{x}\ppr-\vec{x}|}\;
\chi_{\vec{y}}(t_{p})\;\chi_{\vec{y}\ppr}(t_{p})\bigg\}=\int\dtot[s]\;\dtot[\sigma] } \\ \no &&
\exp\bigg\{-\frac{\im}{4\hbar}\int\dtot t_{p}\sum_{\vec{x};\vec{x}\ppr}
\frac{1}{V_{|\vec{x}\ppr-\vec{x}|}}\Tr\bigg[
\Sigma_{\vec{x},s;\vec{x}\ppr,s\ppr}^{ab}(t_{p})\;\kappa^{bb}\;
\Sigma_{\vec{x}\ppr,s\ppr;\vec{x},s}^{ba}(t_{p})\;\kappa^{aa} \bigg]\bigg\} \\ \no &&
\exp\left\{-\frac{\im}{2\hbar}\int_{C}\dtot t_{p}\sum_{\vec{y};\vec{y}\ppr}
\left(
\bea{c}
\chi_{\vec{y}}(t_{p}) \\
\chi_{\vec{y}}^{*}(t_{p})
\eea
\right)^{+}\;\;
\Sigma_{\vec{y};\vec{y}\ppr}(t_{p})\;\;
\left(
\bea{c}
\chi_{\vec{y}\ppr}(t_{p}) \\
\chi_{\vec{y}\ppr}^{*}(t_{p})
\eea
\right)\right\} \\ \no &&
\kappa =\mbox{diag}(1,1,-1,-1)\;\; .
\eeq
Substitution of the interaction with the Hubbard-Stratonovich transformation
yields the following coherent state path integral where a doubling of the
one particle terms has to be performed because of the source terms so that
a bilinear product of anticommuting fields is obtained
\beq \lb{21}
\lefteqn{Z[\mcal{J}]=
\int\dtot[s]\;\dtot[\sigma]\;
\exp\left\{-\frac{\im}{4\hbar}\int_{C}\dtot t_{p}\sum_{\vec{x},\vec{x}\ppr}
\frac{1}{V_{|\vec{x}\ppr-\vec{x}|}}\Tr
\Big[\Sigma\;\kappa\;\Sigma\;\kappa\Big]\right\} } \\ \no &\times &
\int\dtot[\chi]\;
\exp\left\{-\frac{\im}{2\hbar}\int_{C}\dtot t_{p}\;\eta^{+}\left[
\left(
\bea{cc}
\wtilde{\mcal{H}} & -j \\
-j^{+} & -\wtilde{\mcal{H}}^{T}
\eea
\right) +\mcal{J}+\Sigma\right]\;\eta\right\}
\eeq
\beq \lb{22}
j_{ss\ppr}(\vec{x},t_{p}) &=&\left(
\bea{cc}
0 & \widetilde{j}(\vec{x},t_{p}) \\
-\wtilde{j}(\vec{x},t_{p}) & 0
\eea
\right) \\ \lb{23}
\wtilde{\mcal{H}}_{ss\ppr}(t_{p}) &=&\delta_{ss\ppr}\;
\bigg(-\im\hbar\frac{\pp}{\pp t_{p}}-\im\;\ve_{p}+
\frac{\vec{p}^{2}}{2m}+v(\vec{x},t)\bigg)_{.}
\eeq
After integration over the anticommuting variables, the path integral
only contains the self-energy $\Sigma$ with the hermitian matrix $s$ for the density terms and the antisymmetric matrix $\sigma$ for the anomalous terms
\beq \lb{24}
Z[\mcal{J}]&=&
\int\dtot[s]\;\dtot[\sigma]\;
\exp\left\{-\frac{\im}{4\hbar}\int_{C}\dtot t_{p}\sum_{\vec{x},\vec{x}\ppr}
\frac{1}{V_{|\vec{x}\ppr-\vec{x}|}}\Tr
\Big[\Sigma\;\kappa\;\Sigma\;\kappa\Big]\right\} \\ \no &&
\sqrt{\det\left[\left(
\bea{cc}
\wtilde{\mcal{H}}(t_{p}) & -j(t_{p}) \\
-j^{+}(t_{p}) & -\wtilde{\mcal{H}}^{T}(t_{p})
\eea\right) +\mcal{J}+
\left(\bea{cc}
s(t_{p}) & -\sigma(t_{p}) \\
-\sigma^{+}(t_{p}) & -s^{T}(t_{p})
\eea\right)\right]_{.}
}
\eeq
The generating function \(Z[\mcal{J}]\) (\ref{24}) is invariant under a symplectic symmetry
group $Sp(2)$ which has its cause in the anticommuting properties of the fields \(\chi_{\vec{x},s}(t_{p})\) and the doubling of spin space (see appendix \ref{App1}). 
This symplectic invariance is spontaneously broken by the $U(2)$ invariance of 
the matrix $s$ and $-s^{T}$ and the source term \(j(t_{p})\) (\ref{22}) so that three complex or six real Goldstone fields result on the coset space \(Sp(2)\backslash U(2)\) 
because the corresponding dimensions of the Lie algebras for \(Sp(2)\) and \(U(2)\) are ten and four, respectively. 
The densities $s$ and $-s^{T}$ represent the self-energy of the interacting
Fermi sea and can be regarded as {\it a vacuum state or background field}
on which the subgroup $U(2)$ invariantly acts so that the symmetry \(Sp(2)\)
of the complete Lagrangian is spontaneously broken to three complex Goldstone fields.

However, if the trap potential in $v(\vec{x},t)$ can be restricted to a typical distance $a_{0}$ and if this distance $a_{0}$ is considerably larger than the range 
\(r_{0}\approx |\vec{x}\ppr-\vec{x}|\) of the two body potential
\(V_{|\vec{x}\ppr-\vec{x}|}\), there are strong oscillations in the quadratic term
with the nonlocal self-energy \(\Sigma_{\vec{x},s;\vec{x}\ppr,s\ppr}^{ab}(t_{p})\)
of (\ref{24}) because \(1/V_{|\vec{x}\ppr-\vec{x}|}\) tends to infinity
as the short ranged potential \(V_{|\vec{x}\ppr-\vec{x}|}\) approaches zero.
Therefore, the spatially local parts \(\Sigma_{ss\ppr}^{ab}(\vec{x},t_{p})\)
of the self-energy are only retained in the path integral (\ref{24})
\beq \lb{25}
\frac{1}{V_{|\vec{x}\ppr-\vec{x}|}}&\rightarrow&\infty\hspace*{0.5cm}\mbox{for}
\hspace*{0.3cm}|\vec{x}\ppr-\vec{x}|>r_{0} \\  \lb{26}
\Sigma_{\vec{x},s;\vec{x}\ppr,s\ppr}^{ab}(t_{p}) &\rightarrow&
\Sigma_{ss\ppr}^{ab}(\vec{x},t_{p})\;\;\delta_{\vec{x}\ppr,\vec{x}}\;\; .
\eeq
This can be accomplished by integration over the matrix elements
\(\Sigma_{\vec{x},s;\vec{x}\ppr,s\ppr}^{ab}(t_{p})\) with \(|\vec{x}-\vec{x}\ppr|>r_{0}\)
in (\ref{24}) and eliminates the nonlocal parts in the self-energy which cause the
strong oscillation on the nonequilibrium time contour in (\ref{24}).
The diagonal parts \(\sigma_{\vec{x},\uparrow;\vec{x}\ppr,\uparrow}(t_{p})\), 
\(\sigma_{\vec{x},\downarrow;\vec{x}\ppr,\downarrow}(t_{p})\) of spin space in the matrix \(\sigma(t_{p})\) tend to zero
because of the antisymmetry of the pair condensate in the spatial part. The vanishing
of the diagonal elements of spin space in \(\sigma(t_{p})\), due to the assumed short-ranged and spin independent interaction \(V_{|\vec{x}-\vec{x}\ppr|}\), corresponds to the
observation that there is usually no triplett pairing of fermions in the condensate.
Consequently, only one complex Goldstone field for the singlett mode remains whereas
the other two complex fields, which result from spontaneous symmetry breaking \(Sp(2)\backslash U(2)\) in the path integral with nonlocal self-energy,
are suppressed because of the short-ranged interaction.\footnote{In the case of spin dependent forces or other interactions \(V_{|\vec{x}-\vec{x}\ppr|}\) which have their maximum
for \(|\vec{x}-\vec{x}\ppr|\neq 0\), but vanishing zero distance interaction $V_{0}$,
other complex Goldstone fields have to be chosen. These cases are excluded in the
present paper.} A \(Sp(2)\backslash U(2)\) coset space for the Goldstone bosons would result
if the interaction \(V_{|\vec{x}\ppr-\vec{x}|}\) was constant for any pair of spatial
vectors $\vec{x}\ppr$, $\vec{x}$ so that every atom would interact with equal weight with
all other atoms independent of distance.

After a shift of the self-energy matrix by the spontaneous symmetry breaking term,
the coherent state path integral $Z[\mcal{J}]$ (\ref{24}) is transformed to a local
self-energy \(\Sigma_{ss\ppr}^{ab}(\vec{x},t_{p})\) consisting of the hermitian
local matrix $s$ and the antisymmetric local matrix $\sigma$ in spin space and
the two body potential restricted to a finite typical zero distance value $V_{0}<0$
\footnote{The variable $\mcal{N}$ in (\ref{27}) is a normalization factor 
\(\mcal{N}=(L/\Delta x)^{d}\cdot(1/\Delta t)\) because a determinant without 
integration measure is considered in the \(\mbox{Tr}\ln\) term.}
\beq \lb{27}
\lefteqn{Z[\mcal{J}] =\int\dtot[s]\;\dtot[\sigma]} \\ \no &&
\exp\left\{\frac{1}{2}\int_{C}\dtot t_{p}\sum_{\vec{x}}\mcal{N}
\Tr\ln\left[\left(
\bea{cc}
\wtilde{\mcal{H}} & 0 \\
0 & -\wtilde{\mcal{H}}^{T}
\eea\right)+\mcal{J}+\left(
\bea{cc}
s(t_{p}) & -\sigma(t_{p}) \\
-\sigma^{+}(t_{p}) & -s^{T}(t_{p})
\eea\right)\right]\right\} \times \\ \no &\times &
\exp\left\{-\frac{\im}{4\hbar}\int_{C}\dtot t_{p}\sum_{\vec{x}}\frac{1}{V_{0}}
\Tr\left[\left(\Sigma+\left(
\bea{cc}
0 & j \\
j^{+} & 0
\eea
\right)\right)\;\kappa\;
\left(\Sigma+\left(
\bea{cc}
0 & j \\
j^{+} & 0
\eea
\right)\right)\;\kappa\;\right]\right\}_{.}
\eeq
The \(4\times 4\) local self-energy matrix in (\ref{27}) consists of the hermitian
\(U(2)\) invariant density term \(s(t_{p})\) with four real parameters and the
antisymmetric \(2\times 2\) anomalous term with one complex field so that the matrix
\(\Sigma_{ss\ppr}^{ab}(\vec{x},t_{p})\) contains six real fields. The number of
independent parameters and the dimension of \(\Sigma_{ss\ppr}^{ab}(\vec{x},t_{p})\)
indicate a \(SO(4)\) symmetry where the antisymmetry of \(SO(4)\) generators becomes
obvious after exchange of the first, second with the third, fourth row of the 
\(4\times 4\) matrices in (\ref{27}). The following {\it local} parametrization (\ref{34}-\ref{36}) of the self-energy with only one complex field \(\phi(\vec{x},t_{p})\) as Goldstone boson can be chosen in the coset space \(SO(4)\backslash U(2)\) in order to separate the anomalous term from the unchanged interacting Fermi sea with density matrix $s$, $-s^{T}$ in the spontaneous symmetry breaking
\beq \lb{28}
\Sigma_{ss\ppr}^{ab}(\vec{x},t_{p}) &=&\left(
\bea{cc}
s_{ss\ppr}(\vec{x},t_{p}) & \sigma_{ss\ppr}(\vec{x},t_{p}) \\
\sigma_{ss\ppr}^{+}(\vec{x},t_{p}) & -s_{ss\ppr}^{T}(\vec{x},t_{p})
\eea\right) \\ \lb{29}
s_{s\ppr s}^{*}(\vec{x},t_{p})&=&s_{ss\ppr}(\vec{x},t_{p})\hspace*{1.0cm}
\sigma_{s\ppr s}(\vec{x},t_{p})=-\sigma_{ss\ppr}(\vec{x},t_{p}) \\ 
\sigma_{\uparrow\uparrow}(\vec{x},t_{p})&=&\sigma_{\downarrow\downarrow}(\vec{x},t_{p})=0
\eeq
\beq \lb{34}
\Sigma_{ss\ppr}^{ab}(\vec{x},t_{p}) &=& T(\vec{x},t_{p})\;\left(
\bea{cc}
s_{D}(\vec{x},t_{p}) & 0 \\
0 & -s_{D}^{T}(\vec{x},t_{p})
\eea\right)\;T^{+}(\vec{x},t_{p}) \\ \lb{35}
T &=& \left(
\bea{cc}
\sqrt{1+t^{+}t} & t^{+} \\
t & \sqrt{1+tt^{+}}
\eea\right)\hspace*{0.5cm}
s_{D}=\left(
\bea{cc}
u & w \\
w^{*} & v
\eea\right) \\ \lb{36}
t &=&\phi(\vec{x},t_{p})\;1_{2}\hspace*{0.5cm}
w=w_{r}+\im\;w_{i}\hspace*{0.5cm}
u,v,w_{r},w_{i}\in \mbox{\sf R}\;\; .
\eeq
The coherent state path integral can be transformed with the chosen parametrization 
(\ref{34}-\ref{36})
of $\Sigma$ and the change of integration measure \(w_{i}^{2}(\vec{x},t_{p})/4\) to the form
\beq \lb{43}
\lefteqn{Z[\mcal{J}]=\int\dtot[u]\;\dtot[v]\;\dtot[w_{r}]\;\dtot[w_{i}]\;\dtot[\phi]\;
\bigg(\prod_{\{\vec{x},t_{p}\}}\frac{w_{i}^{2}(\vec{x},t_{p})}{4}\bigg) } \\ \no &&
\exp\bigg\{-\frac{\im}{4\hbar}\int_{C}\dtot t_{p}\sum_{\vec{x}}\frac{1}{V_{0}}\bigg(
2\;\trs(s_{D}^{2})-4\;\wtilde{j}\;\wtilde{j}^{*}-16\;w_{i}\;
\Im(\phi\;\wtilde{j})\;\sqrt{1+|\phi|^{2}}\bigg)\bigg\} \\ \no &&
\exp\left\{\frac{1}{2}\int_{C}\dtot t_{p}\sum_{\vec{x}}\mcal{N}\Tr
\ln\left[\left(
\bea{cc}
\wtilde{\mcal{H}} & \\
 & -\wtilde{\mcal{H}}^{T}
 \eea\right)+\mcal{J}+ T\;\left(
 \bea{cc}
 s_{D} & \\
  & -s_{D}^{T}
\eea\right)\;T^{+}\right]\right\}_{.}
\eeq
Since only a small fraction of the Fermi sea condenses, classical equations for the
\(2\times 2\) density matrix \(s_{D}(\vec{x},t_{p})\) can be obtained by variation
of the action in (\ref{43}) with respect to $u$, $v$, $w_{r}$ and $w_{i}$ (\ref{35},\ref{36}) where the matrix $T$ is set to the unit matrix and the integration measure \(w_{i}^{2}(\vec{x},t_{p})\) is included
in the variation\footnote{The fields \(u(\vec{x},t_{p}),\;v(\vec{x},t_{p}),\;w_{r}(\vec{x},t_{p}),\;w_{i}(\vec{x},t_{p})\) 
in \(s_{D}(\vec{x},t_{p})\) have to be scaled by
the factor \((\Delta t/\hbar)\cdot(\Delta x/L)^{d}\) to dimensionless quantities for
the variation.}
\beq\lb{c1}
\lefteqn{\mbox{variation }\;\delta u(\vec{x},t_{p})\;:}  \\ \no && 
-\frac{\im}{V_{0}}\;u^{0}+(\wtilde{\mcal{H}}+s_{D}^{0})_{\uparrow\uparrow}^{-1}(\vec{x},t_{p})= 0 \\ \lb{c2}
\lefteqn{\mbox{variation }\;\delta v(\vec{x},t_{p})\;:}  \\ \no && 
-\frac{\im}{V_{0}}\;v^{0}+(\wtilde{\mcal{H}}+s_{D}^{0})_{\downarrow\downarrow}^{-1}(\vec{x},t_{p})= 0 \\ \lb{c3}
\lefteqn{\mbox{variation }\;\delta w_{r}(\vec{x},t_{p})\;:}  \\ \no && 
-\frac{2\im}{V_{0}}\;w_{r}^{0}+
(\wtilde{\mcal{H}}+s_{D}^{0})_{\uparrow\downarrow}^{-1}(\vec{x},t_{p})+
(\wtilde{\mcal{H}}+s_{D}^{0})_{\downarrow\uparrow}^{-1}(\vec{x},t_{p}) = 0 \\ \lb{c4}
\lefteqn{\mbox{variation }\;\delta w_{i}(\vec{x},t_{p})\;:}  \\ \no && 
-\frac{2\im}{V_{0}}\;w_{i}^{0}+\im\Big[
(\wtilde{\mcal{H}}+s_{D}^{0})_{\uparrow\downarrow}^{-1}(\vec{x},t_{p})-
(\wtilde{\mcal{H}}+s_{D}^{0})_{\downarrow\uparrow}^{-1}(\vec{x},t_{p})\Big]+
\frac{2}{w_{i}^{0}} = 0\;\; .
\eeq
The classical equations (\ref{c1}-\ref{c4}) for the resulting matrix  \(s_{D}^{0}(\vec{x},t_{p})\) 
can be simplified with the Thomas-Fermi approximation where the kinetic energy can be
neglected because of the large atom masses and an assumed homogeneous self-energy of the bulk Fermi sea. This gives algebraic equations for 
\(s_{D}^{0}(\vec{x},t_{p})\) which can be applied in the correlation functions
of the nonlinear sigma model with matrix $T$ (\ref{35}) following in section \ref{s3}.
Using the parametrization into block diagonal densities $s_{D}$ for the Fermi sea and
anomalous terms, the determinant in (\ref{43}) has to be expanded with respect to the
gradients contained in $\wtilde{\mcal{H}}$ \cite{Pruis1}{-}\cite{Pruis3}.

\section{Separation into densities and anomalous terms with a gradient expansion} \lb{s3}

Applying the chosen parametrization (\ref{34}-\ref{36}) for $\Sigma$, we can insert the term
\(\kappa\;T\;\kappa\;T^{+}\;\kappa\) into the \(\mbox{Tr}\ln\) term of (\ref{43}) without
a modification of the coherent state path integral because the determinant of
\(\kappa\) in combined spin and hyperbolic space equals unity
\beq \lb{44}
\lefteqn{\mcal{O}_{1}=\int_{C}\dtot t_{p}\sum_{\vec{x}}\mcal{N}\Tr
\ln\left\{\left[\left(
\bea{cc}
\wtilde{\mcal{H}} & \\
 & -\wtilde{\mcal{H}}^{T}
\eea\right)+\mcal{J}+\Sigma\right]\;\underbrace{\kappa\;T\;\kappa}_{T^{-1}}\;T^{+}\;\kappa
\right\} } \\ \no &=&
\int_{C}\dtot t_{p}\sum_{\vec{x}}\mcal{N}\Tr\ln\Bigg[\left(
\bea{cc}
\wtilde{\mcal{H}}+s_{D} & \\
 & \wtilde{\mcal{H}}^{T}+s_{D}^{T}
 \eea\right)+T\;\kappa\;\mcal{J}\;T^{-1}+ \\ \no &+&
 \underbrace{T\;\left(
 \bea{cc}
 \wtilde{\mcal{H}} & \\
  & \wtilde{\mcal{H}}^{T}
  \eea\right)\;T^{-1}-\left(
  \bea{cc}
  \wtilde{\mcal{H}} & \\
   & \wtilde{\mcal{H}}^{T}
\eea\right)}_{\delta\mcal{H}_{ss\ppr}^{ab}}\Bigg]_{.}
\eeq
The gradient expansion of \(\delta\mcal{H}_{ss\ppr}^{ab}\) gives the following
operator \(\delta\mcal{H}_{ss\ppr}^{ab}=\delta_{ss\ppr}\;\delta\mcal{H}^{ab}\)
\beq \lb{45}
\lefteqn{
T\;\left(
\bea{cc}
\wtilde{\mcal{H}} & \\
 & \wtilde{\mcal{H}}^{T}
\eea\right)\;T^{-1}-\left(
\bea{cc}
\wtilde{\mcal{H}} & \\
  & \wtilde{\mcal{H}}^{T}
\eea\right)= } \\ \no &=&
(T\;\kappa\;T^{-1}-\kappa)\;(-\hat{E}_{p})+
T\;\kappa\;(-E_{p}T^{-1})  \\ \no  &-&
\frac{\hbar^{2}}{2m}(\pp_{\mu}T)\;T^{-1}\;(\pp_{\mu}T)\;T^{-1}+
\frac{\hbar^{2}}{2m}(\pp_{\mu}T)\;T^{-1}\;\hat{\pp}_{\mu}+
\frac{\hbar^{2}}{2m}\hat{\pp}_{\mu}\;(\pp_{\mu}T)\;T^{-1} \\ \lb{46} &&
\hat{E}_{p}=\im\hbar\;\frac{\hat{\pp}}{\pp t_{p}}_{,}
\eeq
where one has to distinguish between the operators \(\hat{\pp}_{\mu}\),
\(\hat{E}_{p}\) and the derivatives
\((\pp_{\mu}T)\),\((E_{p}T)=\im\hbar(\pp T/\pp t_{p})\) of the matrices \(T,\;T^{-1}\), e.g.
\(T\;\hat{\pp}_{\mu}\;T^{-1}=T\;\Big[(\pp_{\mu}T^{-1})+T^{-1}\;\hat{\pp}_{\mu}\Big]\).

Expanding the \(\mbox{Tr}\ln\) term $\mcal{O}_{1}$ up to second order in
\(\delta\mcal{H}_{ss\ppr}^{ab}\), we obtain the expression (\ref{47})
with the Green function \(G_{ss\ppr}^{a}\) of the block diagonal density matrix $s_{D}$ in  spin space in symbolic form (spatial and time coordinates are omitted for brevity). In the following the field $w_{i}$
in the matrix $s_{D}$ (\ref{35},\ref{36}) for the Green function \(G_{ss\ppr}^{a}\)
has to be separated from the expansion because it couples to the 
orignal $U(1)$ symmetry violating source term $\wtilde{j}$ (\ref{43},\ref{6})
\beq \lb{47}
\mcal{O}_{1}&=&2\int_{C}\dtot t_{p}\sum_{\vec{x}}\mcal{N}\trs\ln\Big(
\wtilde{\mcal{H}}+s_{D}\Big)
\\ \no &+& \int_{C}\dtot t_{p}\sum_{\vec{x}}\mcal{N}\Tr
\Big[G_{ss\ppr}^{a}\;(T\;\kappa\;\mcal{J}\;T^{-1}+\delta\mcal{H})_{s\ppr s}^{aa}\Big] \\ \no
&-&\frac{1}{2}\int_{C}\dtot t_{p}\sum_{\vec{x}}\mcal{N}\Tr\Big[
G_{ss\ppr}^{a}\;\delta\mcal{H}^{ab}\;G_{s\ppr s}^{b}\;\delta\mcal{H}^{ba}\Big] \\ \no &-&
\frac{1}{2}\int_{C}\dtot t_{p}\sum_{\vec{x}}\mcal{N}\Tr\Big[
G_{s_{1}s_{1}\ppr}^{a_{1}}\;(T\;\kappa\;\mcal{J}\;T^{-1})_{s_{1}\ppr s_{2}\ppr}^{a_{1}a_{2}}\;
G_{s_{2}\ppr s_{2}}^{a_{2}}\;(T\;\kappa\;\mcal{J}\;T^{-1})_{s_{2}s_{1}}^{a_{2}a_{1}}\Big] \\ \no &-&
\int_{C}\dtot t_{p}\sum_{\vec{x}}\mcal{N}\Tr\Big[
G_{s_{1}s_{1}\ppr}^{a_{1}}\;(T\;\kappa\;\mcal{J}\;T^{-1})_{s_{1}\ppr s_{2}\ppr}^{a_{1}a_{2}}\;
G_{s_{2}\ppr s_{1}}^{a_{2}}\;
\delta\mcal{H}^{a_{2}a_{1}}\Big]_{.}
\eeq
The hermitian self-energy $s_{D;ss\ppr}(\vec{x},t_{p})$ 
without the field $w_{i}$
is only contained in the Green function \(G_{ss\ppr}^{a}(\vec{x},t_{p};\vec{x}\ppr,t_{p}\ppr)\) on which the
derivative operators in \(\delta\mcal{H}^{ab}\) act.  
A spatial and time diagonal Green function \(g_{ss\ppr}(\vec{x},t_{p})\) for \(G_{ss\ppr}^{a}\) can be considered in a Thomas-Fermi approximation
for large atom masses and a nearly homogeneous system where the kinetic energy and time derivative can be added as a perturbation 
\beq \lb{48}
\lefteqn{
G_{ss\ppr}^{a(=1/2)}(\vec{x},t_{p};\vec{x}\ppr,t_{p}\ppr) =
\langle\vec{x},s;t_{p}|(\wtilde{\mcal{H}}+s_{D})^{(T)-1}|\vec{x}\ppr,s\ppr;t_{p}\ppr\rangle } \\ \no
&=& \underbrace{\Big(-\im\ve_{p}+v(\vec{x},t_{p})+s_{D}(\vec{x},t_{p})\Big)_{ss\ppr}^{-1}}_{
g_{ss\ppr}(\vec{x},t_{p})}\;
\delta_{\vec{x},\vec{x}\ppr}\;\delta(t_{p}-t_{p}\ppr)+\mbox{derivative terms}\; .
\eeq
The first order term of \(\delta \mcal{H}_{ss\ppr}^{ab}\) in $\mcal{O}_{1}$ 
(second term in \ref{47}) vanishes in the Thomas-Fermi approximation with \(G_{ss\ppr}^{a}\) replaced by the spatial and time diagonal function \(g_{ss\ppr}(\vec{x},t_{p})\) (\ref{48}). The anomalous terms with the matrix $T$ can be reduced to a \(2\times 2\) matrix because the spin space is restricted to the block diagonal densities. 

Introducing the following averages of the block diagonal densities with matrix 
\(s_{D;ss\ppr}(\vec{x},t_{p})\)
\beq \lb{49}
\Big\langle\ldots\Big\rangle &=&\int\dtot[s_{D}]\;
\prod_{\{\vec{x},t_{p}\}}\frac{w_{i}^{2}(\vec{x},t_{p})}{4}\;\Big(\ldots\Big)\;\;
\det\Big(\mcal{H}+s_{D}\Big) \\ \no &\times &
\exp\bigg\{-\frac{\im}{2\hbar}\frac{1}{V_{0}}\int_{C}\dtot t_{p}\sum_{\vec{x}}
\;\trs\Big(s_{D}(\vec{x},t_{p})\Big)^{2}\bigg\}_{.}
\eeq
\beq \lb{50}
\frac{\im}{\hbar V_{0}}\;c_{tt}(\vec{x},t_{p}) &=&
\Big\langle\trs\Big[ (E_{p}g_{ss\ppr}(\vec{x},t_{p}))\;
(E_{p}g_{s\ppr s}(\vec{x},t_{p}))\Big]\Big\rangle \\ \lb{51}
\frac{\im}{\hbar V_{0}}\;c_{t}(\vec{x},t_{p}) &=&
\Big\langle\trs\Big[
(-E_{p}g_{ss\ppr}(\vec{x},t_{p}))\;g_{s\ppr s}(\vec{x},t_{p})\Big]\Big\rangle \\ \lb{52}
\frac{\im}{\hbar V_{0}}\;c_{\mu t}(\vec{x},t_{p}) &=&
\Big\langle\trs\Big[
(\pp_{\mu}g_{ss\ppr}(\vec{x},t_{p}))\;
(-E_{p}g_{s\ppr s}(\vec{x},t_{p}))\Big]\Big\rangle \\ \lb{53}
\frac{\im}{\hbar V_{0}}\;c(\vec{x},t_{p}) &=&
\Big\langle\trs\Big[
g_{ss\ppr}(\vec{x},t_{p})\;g_{s\ppr s}(\vec{x},t_{p})\Big]\Big\rangle \\ \lb{54}
\frac{\im}{\hbar V_{0}}\;c_{\mu}(\vec{x},t_{p}) &=&
\Big\langle\trs\Big[
g_{ss\ppr}(\vec{x},t_{p})\;
(\pp_{\mu}g_{s\ppr s}(\vec{x},t_{p}))\Big]\Big\rangle \\ \lb{55}
\frac{\im}{\hbar V_{0}}\;c_{\mu\nu}(\vec{x},t_{p}) &=&
\Big\langle\trs\Big[
(\pp_{\mu}g_{ss\ppr}(\vec{x},t_{p}))\;
(\pp_{\nu}g_{s\ppr s}(\vec{x},t_{p}))\Big]\Big\rangle\;\; ,
\eeq
the coherent state path integral can be separated into an action \(\mcal{S}_{0}[s_{D}]\)
following from (\ref{49}) and an action \(\mcal{S}[T,T^{-1};\{c_{ij}\}]\) for the anomalous terms
\beq \lb{56}
Z[\mcal{J}] &\approx&\int\dtot[s_{D}]\;\bigg(\prod_{\{\vec{x},t_{p}\}}
\frac{w_{i}^{2}(\vec{x},t_{p})}{4}\bigg)\;
\exp\Big\{-\im\;\mcal{S}_{0}[s_{D}]\Big\} \\ \no &\times & \int\dtot[\phi]\;
\exp\Big\{-\im\;\mcal{S}[T,T^{-1};\{c_{ij}\}]\Big\}\;\;
\exp\Big\{-\im\;\mcal{S}_{\mcal{J},j}[T,s_{D};\mcal{J},j]\Big\}_{.}
\eeq
The action \(\mcal{S}[T,T^{-1};\{c_{ij}\}]\) is given by the following relation up to second
order in the gradients $\hat{\pp}_{\mu}$, $\hat{E}_{p}$ where partial spatial integrations have been performed. The parameter functions \(c_{ij}(\vec{x},t_{p})\) (\ref{50}-\ref{55})
contain the properties of the densities with the matrix $s_{D}$ as a background field
\beq \lb{57a}
T^{-1}&=&\kappa\;T\;\kappa \hspace*{1.0cm}\kappa=\mbox{diag}(1,-1) \\ \lb{57}
(\mcal{D}T^{-1})&=&-(E_{p}T^{-1})=\im\hbar\;T^{-1}\;(\pp_{t_{p}}T)\;T^{-1}
\eeq
\beq \lb{58}
\lefteqn{S[T,T^{-1};\{c_{ij}\}]=\frac{1}{\hbar V_{0}}\int_{C}\dtot t_{p}\sum_{\vec{x}}
} \\ \no && \Bigg\{ c(\vec{x},t_{p})\;\trab\Big[(T\kappa(\mcal{D}T^{-1}))^{2}\Big]-
c_{\mu\nu}(\vec{x},t_{p})\;\bigg(\frac{\hbar^{2}}{m}\bigg)^{2}
\trab\Big[(\pp_{\mu}T)\;(\pp_{\nu}T^{-1})\Big]
\\ \no &+& c_{tt}(\vec{x},t_{p})\;\trab\Big[(T\kappa T^{-1}-\kappa)^{2}\Big] +
c_{\mu}(\vec{x},t_{p})\;\frac{2\hbar^{2}}{m}\trab\Big[(\pp_{\mu}T)\kappa(\mcal{D}T^{-1})\Big]
\\ \no &+&
\Big(2\;c_{\mu t}(\vec{x},t_{p})-\pp_{\mu}c_{t}(\vec{x},t_{p})\Big)\;\frac{\hbar^{2}}{m}\;
\trab\Big[[T,\kappa]\;(\pp_{\mu}T)\Big]
\\ \no &+& 2\;c_{t}(\vec{x},t_{p})\;\trab\Big[(T^{-1}-T)\;(\mcal{D}T)\Big]\Bigg\}_{.} 
\eeq
It can be verified with an expansion of the Goldstone field \(\phi(\vec{x},t_{p})\) of the
matrix $T$ in \(\sinh\)-amplitude and phase term \(\phi(\vec{x},t_{p})=
\sinh(\varphi(\vec{x},t_{p}))\;\exp\{\im\;\alpha(\vec{x},t_{p})\}\) (\(\varphi(\vec{x},t_{p})\geq 0\)) that the derived action (\ref{58}) is composed
of a massless Goldstone field $\alpha(\vec{x},t_{p})$ and a real massive field
\(\sinh(\varphi(\vec{x},t_{p}))\). The parameter functions \(c_{ij}(\vec{x},t_{p})\) can be 
regarded as generalized mass and kinetic terms of a spontaneously broken $\phi^{4}$ field theory\cite{Hua}.  They can be calculated from the classical equations (\ref{c1}-\ref{c4})
for \(s_{D}^{0}(\vec{x},t_{p})\) where the Thomas-Fermi approximation of large atom mass
can be used for simplicity. In terms of the $\sinh$-amplitude and phase term, the action
\(S[T,T^{-1};\{c_{ij}\}]\) (\ref{58}) is similar to that of generalized Sine-Gordon equations
\beq \lb{60}
\lefteqn{S[\varphi,\alpha;\{c_{ij}\}]=\frac{1}{\hbar V_{0}}\int_{C}\dtot t_{p}\sum_{x}}
\\ \no && \Bigg\{2\hbar^{2}\;\sinh^{2}(\varphi)\;\bigg[c(\vec{x},t_{p})\;\cosh(2\varphi)\;
(\pp_{t_{p}}\alpha)^{2}+c_{\mu\nu}(\vec{x},t_{p}) \bigg(\frac{\hbar}{m}\bigg)^{2}\;(\pp_{\mu}\alpha)\;(\pp_{\nu}\alpha)\bigg]
\\ \no &+& 2\hbar^{2}\bigg[
c(\vec{x},t_{p})\;(\pp_{t_{p}}\varphi)^{2}+c_{\mu\nu}(\vec{x},t_{p})
\bigg(\frac{\hbar}{m}\bigg)^{2}\;
(\pp_{\mu}\varphi)\;(\pp_{\nu}\varphi)\bigg]-8\;c_{tt}(\vec{x},t_{p})\;\sinh^{2}(\varphi) 
\\ \no &+&\frac{2\hbar^{3}}{m}\;c_{\mu}(\vec{x},t_{p})\;\sinh(2\varphi)\;\;
\Big[(\pp_{t_{p}}\alpha)\;(\pp_{\mu}\varphi)-(\pp_{\mu}\alpha)\;(\pp_{t_{p}}\varphi)\Big]
\\ \no &-&
\frac{4\im\hbar^{2}}{m}\Big(2\;c_{\mu t}(\vec{x},t_{p})-\pp_{\mu}c_{t}(\vec{x},t_{p})\Big)\;
\sinh^{2}(\varphi)\;(\pp_{\mu}\alpha)+4\im\hbar\;c_{t}(\vec{x},t_{p})\;\sinh(2\varphi)\;
(\pp_{t_{p}}\varphi)
\Bigg\}_{.}
\eeq
A first order variation of the action (\ref{60}) with respect to the fields on the
time contour gives a classical equation for the time development of the Goldstone
field \(\overline{\phi}(\vec{x},t)=\sinh(\bvarphi(\vec{x},t))\times\)
\(\exp\{\im\;\balpha(\vec{x},t)\}\) with the densities $s_{D;ss\ppr}(\vec{x},t_{p})$ as background fields in the parameter functions $\overline{c}_{ij}(\vec{x},t)$ 
\beq \lb{61}
\overline{\varphi}(\vec{x},t) &=&\frac{1}{2}\Big(\varphi(\vec{x},t_{+})+
\varphi(\vec{x},t_{-})\Big) \\ \lb{62}
\overline{\alpha}(\vec{x},t) &=&\frac{1}{2}\Big(\alpha(\vec{x},t_{+})+
\alpha(\vec{x},t_{-})\Big) \\ \lb{63}
\overline{c}_{ij}(\vec{x},t) &=&\frac{1}{2}\Big(c_{ij}(\vec{x},t_{+})+
c_{ij}(\vec{x},t_{-})\Big)
\eeq
\beq\lb{64}
\lefteqn{\pp_{t}\Big(\ovv{c}\;\sinh^{2}(\bvarphi)\;\cosh(2\bvarphi)\;(\pp_{t}\balpha)\Big)+
\bigg(\frac{\hbar}{m}\bigg)^{2}\pp_{\mu}\Big(\ovv{c}_{\mu\nu}\;\sinh^{2}(\bvarphi)\;
(\pp_{\nu}\balpha)\Big)+ } \\ \no &+&\frac{\hbar}{2m}\;\sinh(2\bvarphi)\Big(
(\pp_{t}\ovv{c}_{\mu})\;(\pp_{\mu}\bvarphi)-(\pp_{\mu}\ovv{c}_{\mu})\;(\pp_{t}\bvarphi)\Big)
-\frac{\im}{m}\pp_{\mu}\Big((2\ovv{c}_{\mu t}-(\pp_{\mu}\ovv{c}_{t}))\;\sinh^{2}(\bvarphi)
\Big)=0
\eeq
\beq\lb{65}
\lefteqn{\pp_{t}\Big(\ovv{c}\;(\pp_{t}\bvarphi)\Big)+\bigg(\frac{\hbar}{m}\bigg)^{2}
\;\pp_{\mu}\Big(\ovv{c}_{\mu\nu}\;(\pp_{\nu}\bvarphi)\Big)= } \\ \no
&&\sinh(2\bvarphi)\times\bigg\{\ovv{c}\Big(\cosh(2\bvarphi)-
\frac{1}{2}\Big)(\pp_{t}\balpha)^{2}+
\bigg(\frac{\hbar}{m}\bigg)^{2}\frac{\ovv{c}_{\mu\nu}}{2}(\pp_{\mu}\balpha)(\pp_{\nu}\balpha)
-\frac{2}{\hbar^{2}}\ovv{c}_{tt}-\frac{\im}{\hbar}(\pp_{t}\ovv{c}_{t})+ \\ \no &-&
\frac{\hbar}{2m}(\pp_{\mu}\ovv{c}_{\mu})(\pp_{t}\balpha)+\Big(\frac{\hbar}{2m}
(\pp_{t}\ovv{c}_{\mu})-\frac{\im}{m}(2\ovv{c}_{\mu t}-\pp_{\mu}\ovv{c}_{t})\Big)\;
(\pp_{\mu}\balpha)\bigg\}_{.}
\eeq
The equations (\ref{64},\ref{65}) determine the time development of the Goldstone field
for the condensate composed of fermionic atoms and are analogous to the Gross-Pitaevskii
equation for bosonic constituents.

\section{Summary and discussion}

Since the parameters $c_{ij}(\vec{x},t_{p})$, following from the densities in the
background, change slowly in time and spatial coordinates, the pair condensate is
determined by the lowest order terms of the nonlinear sigma model with the matrix $T$
\be \lb{59}
T=\left(
\bea{cc}\sqrt{1+|\phi|^{2}} & \phi^{*} \\
\phi & \sqrt{1+|\phi|^{2}}
\eea\right)_{.}
\ee
This separation into background properties with the parameters $c_{ij}$ and anomalous
terms of the nonlinear sigma model corresponds to the observation that only a small
fraction of the atoms condense in the Fermi sea. In the case of computations one
therefore can introduce correlation functions of the densities, as e.g. the
current-current correlation function \(c_{\mu\nu}(\vec{x},t_{p})\), etc. which can
be calculated with the self-energy matrix \(s_{D}^{0}(\vec{x},t_{p})\) (\ref{c1}-\ref{c4})
of the first order variation of the action in (\ref{43}).
Taking derivatives of the action \(\mcal{S}_{\mcal{J},j}[T,s_{D};\mcal{J},j]\)
with respect to $\mcal{J}$, the appropriate observables can be obtained, as e.g.
the anomalous term \(\frac{-\im}{\hbar}\langle\chi_{\vec{x},\uparrow}(t_{p})\;
\chi_{\vec{x}\ppr,\downarrow}(t_{p}\ppr)\rangle\) of two anticommuting fields is
represented by a relation with the bosonic matrix $T$ and the Green function
\(G_{ss\ppr}^{a}\) of the densities
\beq\lb{e}
\lefteqn{-\frac{\im}{\hbar\;\mcal{N}}\;\langle\chi_{\vec{x},\uparrow}(t_{p})\;
\chi_{\vec{x}\ppr,\downarrow}(t_{p}\ppr)\rangle = } \\ \no &=&\sum_{a=1}^{2}
(T^{-1})^{2a}(\vec{x}\ppr,t_{p}\ppr)\;\Big[G_{\downarrow\uparrow}^{a}(\vec{x}\ppr,t_{p}\ppr;
\vec{x},t_{p})-G_{\uparrow\downarrow}^{a}(\vec{x},t_{p};\vec{x}\ppr,t_{p}\ppr)\Big]\;
T^{a1}(\vec{x},t_{p})\;\;.
\eeq 
Statements about binding energies of the pairs \(\langle\chi_{\vec{x},\uparrow}(t_{p})\;
\chi_{\vec{x},\downarrow}(t_{p})\rangle\) can be calculated by various saddle point
considerations from the coherent state path integral (\ref{24}) \cite{E1,E2}. However,
the exact expression of the saddle point is not needed for the derivation of the
nonlinear sigma model with actions (\ref{58}) and (\ref{60}) for the time development of a condensate composed of fermionic atoms.
The equations (\ref{64}) and (\ref{65}) simplify considerably for a translation invariant
system with a spatially constant and time independent external potential $u(\vec{x},t)$
(\ref{5}) so that only terms with the constant coefficients \(\ovv{c}\) (\ref{53}) and \(\ovv{c}_{\mu\nu}\) (\ref{55}) remain. In this case a simpler form of the Sine-Gordon equation results for the condensate of the Goldstone field $\phi$ in the matrix $T$ (\ref{59}).

In experiments fermionic $^{6}\mbox{Li}$ was cooled to degeneracy by $^{23}\mbox{Na}$ or 
by mixing two different states in an optical trap \cite{Ha,Gra} and fermionic $^{40}\mbox{K}$ was cooled to \(T/T_{F}=0.3,\;(E_{F}=k_{B}T_{F})\) by sympathetic cooling 
with \(^{87}\mbox{Rb}\) \cite{Roa}. Apart from degeneracy a BCS transition, for which the
time development of the condensate is described by Eqs. (\ref{49}) to (\ref{65}),
is suggested in Ref. \cite{Schr}. For this a degenerate Fermi gas with
attractive interaction between the fermions must be prepared. A possible realization can 
be obtained in fermionic $^{6}\mbox{Li}$ \cite{Schr}. In experiments atoms can be prepared
in two different hyperfine states and a Feshbach resonance is used to tune the interaction
due to elastic collisions to the desired value \cite{Tie}. In Refs. \cite{Com}-\cite{Tim}, 
BCS transition temperatures $T_{BCS}$ are predicted in the range from
\(T_{BCS}/T_{F}=0.025\) to \(T_{BCS}/T_{F}=0.4\), but it is an open question
if these conditions can be reached experimentally.

Close to a Feshbach resonance, the additional phenomenon of a BCS to BEC crossover
is predicted for atomic Fermi gases \cite{Oha}. A strong attractive interaction can arise
between fermions mediated by bosonic quasi-molecules associated with a Feshbach resonance.
In this case it is stated that one has to be careful in applying pure BCS theory to a
Fermi gas when the pairing interaction is very strong \cite{Oha}, due to fluctuations
in the two-particle Cooper channel.
One can also try to extend the experiments in optical lattices with bosonic constituents
to fermionic atoms \cite{Bl1}. However, optical lattices involve the additional length scale of the periodic potential so that one has to examine whether a gradient expansion up to
second order is sufficient for describing the atoms interacting on one lattice site.
In this case one has to take into account the band structure of the optical lattice
in the expansion of the \(\mbox{Tr ln}\)-term of relation (\ref{44}).

In the present paper we have performed symmetry considerations on which the derivation
of the nonlinear sigma model with the Goldstone field $\phi$ in the coset matrix $T$
(\ref{59}) is based. Since the Sine-Gordon equation allows nontrivial solitonic solutions
in 1+1 or two spatial dimensions, one can also expect solitons in condensates composed
of fermions and study their temporal evolution. Apart from numerical computations,
the derived effective Gross-Pitaevskii equation for fermions gives rise to investigations
for B\"acklund-transformations in 1+1 or two spatial dimensions so that nontrivial
solutions can be obtained as for the nonlinear Schr\"odinger equation of
bosonic condensates.
\vspace*{1.0cm}

\noindent {\bf Acknowledgement}\newline
\noindent This work was supported by the DFG within the research program SFB/TR12
"{\it Symmetries and Universality in Mesoscopic Systems}". 

\newpage

\begin{appendix} 

\section{Symplectic symmetry of the generating function}\lb{App1}

In the following we abbreviate spatial and time coordinates \(\vec{x}\), \(t_{p}\)
of the fields \(\chi_{\vec{x},s}(t_{p})\), \(\chi_{\vec{x},s}^{*}(t_{p})\) with
the indices $i,\;j$ in the global transformations acting in the doubled spin space.
The symmetries result from the spin independent one particle Hamiltonian $\wtilde{\mcal{H}}$
and the spin independent interaction. Considering the Hamiltonian \(\wtilde{\mcal{H}}_{ij}\)
as a spin independent matrix, we can rewrite the bilinear Hamiltonian form with
the antisymmetry of the fields \(\chi_{i,s}\), \(\chi_{i,s}^{*}\) as the relation
\beq \lb{b1}
\chi_{i,s}^{*}\;\wtilde{\mcal{H}}_{ij}\;\chi_{j,s} &=&\frac{1}{2}
\Big(\chi_{i,s}^{*}\;\wtilde{\mcal{H}}_{ij}\;\chi_{j,s}-\chi_{j,s}\;
\wtilde{\mcal{H}}_{ij}\;\chi_{i,s}^{*}\Big) \\ \no &=&\frac{1}{2}
\left(
\bea{c}
\chi_{i,s}^{*} \\ \chi_{i,s}
\eea\right)^{T}\left(
\bea{cc}
\wtilde{\mcal{H}}_{ij} & 0 \\
0 & -\wtilde{\mcal{H}}_{ij}^{T}
\eea\right)\left(
\bea{c}
\chi_{j,s} \\ \chi_{j,s}^{*}
\eea\right)_{.}
\eeq
After a transformation with a matrix $M$, one has to obtain again a separation into
fields \(\chi_{i,s}\ppr\) and their complex conjugates \(\chi_{i,s}^{*\prime}\)
in the first, second and third, fourth row. Therefore, the global transformations
with the \(4\times 4\) matrix $M$ consist of two \(2\times 2\) block matrices $A$, $B$
in spin space and their complex conjugates
\be \lb{b2}
\underbrace{\left(
\bea{c}
\chi_{i,s\ppr}\ppr \\ \chi_{i,s\ppr}^{\prime*}
\eea\right)}_{\eta_{i}\ppr}=\underbrace{\left(
\bea{cc}
A_{s\ppr s} & B_{s\ppr s} \\
B_{s\ppr s}^{*} & A_{s\ppr s}^{*}
\eea\right)}_{M}\underbrace{\left(
\bea{c}
\chi_{i,s} \\ \chi_{i,s}^{*}
\eea\right)}_{\eta_{i}} \;\; .
\ee
Because of relation (\ref{b1}) the matrix $M$ has to be invariant under the following
unitary hyperbolic transformation
\be \lb{b3}
M^{+}\;\kappa\;M=\kappa\hspace{1.0cm}
\kappa=\left(
\bea{cc}
1_{2} & \\
 & -1_{2}
\eea\right)
\ee
or the equivalent symplectic form which results from interchanging in $\kappa$ the first and second row with the third and fourth row, respectively, and from reordering the matrix $M^{+}$
to its transpose $M^{T}$ 
\be \lb{b4}
M^{T}\;g\;M=g\hspace{1.0cm}
g=\left(
\bea{cc}
0 & -1_{2} \\
1_{2} & 0
\eea\right)_{.}
\ee
The bilinear form \(\chi_{i,\uparrow}^{*}\chi_{i,\uparrow}+\chi_{i,\downarrow}^{*}
\chi_{i,\downarrow}\) is invariant under these transformations with the matrix $M$,
both for the unitary hyperbolic and symplectic kind of transformation
\beq \lb{b5}
2\;(\chi_{i,\uparrow}^{*}\chi_{i,\uparrow}+\chi_{i,\downarrow}^{*}
\chi_{i,\downarrow}) &=& \eta_{i}^{+}\;\kappa\;\eta_{i}=
\eta_{i}^{T}\;g\;\eta_{i} \\ \no &=&
\eta_{i}^{\prime+}\;\kappa\;\eta_{i}\ppr=
\eta_{i}^{\prime T}\;g\;\eta_{i}\ppr \\ \no &=&
2\;(\chi_{i,\uparrow}^{\prime*}\chi_{i,\uparrow}\ppr+\chi_{i,\downarrow}^{\prime*}
\chi_{i,\downarrow}\ppr)\;\;.
\eeq
The complex \(2\times 2\) matrices $A$, $B$ consist of sixteen real parameters which are
restricted by the equations (\ref{b3}) or (\ref{b4}), yielding six real conditional
relations. Therefore, ten independent parameters remain in the matrices $A$, $B$.
This indicates a symplectic symmetry group $Sp(N/2)$ with \(\frac{1}{2}N(N+1)\)
parameters for \(N=4\) of the \(4\times 4\) global transformation matrix $M$.
The matrix $M$ can be expressed with the global $U(2)$ subgroup in spin space and the
coset space \(Sp(2)\backslash U(2)\) which is composed of the complex
symmetric matrices $t_{ss\ppr}$ or $t\ppr_{ss\ppr}$ with six and ten real
parameters, respectively
\beq \lb{b6}
M&=&\left(
\bea{cc}
U^{+} & \\
 & U^{T}
\eea\right)\left(
\bea{cc}
\sqrt{1+t^{+}\;t} & t^{+} \\
t & \sqrt{1+t\;t^{+}}
\eea\right)\left(
\bea{cc}
U & \\
 & U^{*}
\eea\right) \\ \no &=&
\left(
\bea{cc}
\sqrt{1+t^{\prime+}\;t\ppr} & t^{\prime+} \\ \lb{b7}
t\ppr & \sqrt{1+t\ppr\; t^{\prime+}}
\eea\right) \\ 
t\ppr &=& U^{T}\;t\;U \hspace*{1.0cm}t=t^{T}\hspace*{0.5cm}t\ppr=t^{\prime T}\; .
\eeq
The matrices $A$, $B$ and their complex conjugates are therefore given by the
following equations which contain ten real independent parameters
\beq \lb{b8}
A&=&\sqrt{1+t^{\prime+}\;t\ppr} \hspace*{1.0cm}B=t^{\prime+} \\ \lb{b9}
A^{*}&=&\sqrt{1+t\ppr\; t^{\prime+}} \hspace*{1.0cm} B^{*}=t\ppr\;\; .
\eeq

\end{appendix}


\begin{thebibliography}{10}
\bibitem{Pit} L. Pitaevskii and S. Stringari, {\it Bose-Einstein Condensation}
(Oxford University Press, Oxford, 2003)
\bibitem{Peth} C.J. Pethick and H. Smith, {\it Bose-Einstein Condensation in Dilute Gases}
(Cambridge University Press, Cambridge, 2002)
\bibitem{Klau} J.R. Klauder and B.S. Skagerstam, {\it Coherent States (Applications in
Physics and Mathematical Physics)} (World Scientific, Singapore, 1985)
\bibitem{Cau} V. Caudrelier and E. Ragoucy, {\it J. Math. Phys.} {\bf 44}, 5706 (2003)
\bibitem{Pruis1} A.M.M. Pruisken, {\it Nucl. Phys.} {\bf B235[FS11]}, 277-298 (1984)
\bibitem{Pruis2} A.M.M. Pruisken, {\it Nucl. Phys.} {\bf B240[FS12]}, 30-48 (1984)
\bibitem{Pruis3} A.M.M. Pruisken, {\it Nucl. Phys.} {\bf B240[FS12]}, 49-70 (1984)
\bibitem{Ball} L.E. Ballentine, {\it Quantum Mechanics}, (Chap. 10), (World Scientific,
Singapore, 1998)
\bibitem{Mosk} S.A. Moskalenko and D.W. Snoke, {\it Bose-Einstein Condensation of Excitons and Biexcitons (and Coherent Nonlinear Optics with Excitons)}, (Cambridge University Press,
Cambridge, 2000)
\bibitem{L1} I.V. Lerner, {\it Nonlinear Sigma Model for Normal and Superconducting Systems: A Pedestrian Approach} in "Proceedings of the International School of Physics" (Enrico Fermi) Course CLI, edited by B. Altshuler and V. Tognetti, (IOS Press, Amsterdam 2003)
\bibitem{L2} I.V. Yurkevich and I.V. Lerner, {\it Phys. Rev.} {\bf B63}, 064522 (2001)
\bibitem{Sch} J. Schwinger, {\it J. Math. Phys.} {\bf 2}, 407 (1961)
\bibitem{Ba1} P.M. Bakshi and K.T. Mahanthappa, {\it J. Math. Phys.} {\bf 4}, 1 (1963)
\bibitem{Ba2} P.M. Bakshi and K.T. Mahanthappa, {\it J. Math. Phys.} {\bf 4}, 12 (1963)
\bibitem{Ke} L.P. Keldysh, {\it Sov. Phys. JETP} {\bf 20}, 1018 (1965)
\bibitem{Neg} J.W. Negele and H. Orland, {\it Quantum Many-Particle Systems},
(Addison-Wesley, Reading, MA, 1988)
\bibitem{Gil1} W.M. Zhang, D.H. Feng and R. Gilmore, {\it Coherent states: theory and some
applications}, {\it Rev. Mod. Phys.} {\bf 62}(4), 867 (1990)
\bibitem{Kl} H. Kleinert, {\it Path Integrals in Quantum Mechanics, Statistics and
Polymer Physics}, (World Scientific, Singapore, 1990)
\bibitem{Ka} T. Kashiwa, Y. Ohnuki and M. Suzuki, {\it Path Integral Methods},
(Oxford Science Publications, Clarendon Press, Oxford 1997)
\bibitem{Gold} J. Goldstone, {\it Nuovo Cimento} {\bf 19}, 154 (1961)
\bibitem{Nambu} Y. Nambu, {\it Phys. Rev. Lett.} {\bf 4}, 380 (1960)
\bibitem{Wen} Xiao-Gang Wen, {\it Quantum Field Theory of Many-Body Systems},
(Oxford University Press, Oxford, 2004)
\bibitem{St} R.L. Stratonovich, {\it Sov. Phys. Dokl.} {\bf 2}, 416 (1958)
\bibitem{Hua} K. Huang, {\it Quarks, Leptons and Gauge Fields}, (World Scientific, Singapore,
1992)
\bibitem{E1} C.A.R. S\'a de Melo, M. Randeria and J.R. Engelbrecht, {\it Phys. Rev. Lett.}
{\bf 71}, 3202 (1993)
\bibitem{E2} J.R. Engelbrecht, M. Randeria and C.A.R. S\'a de Melo,
{\it Phys. Rev. B} {\bf 55}, 15153 (1997)
\bibitem{Ha} Z. Hadzibabic, C.A. Stan, K. Dieckmann, S. Gupta, M.W. Zwierlein, A. G\"{o}rlitz,
W. Ketterle, {\it Phys. Rev. Lett.} {\bf 88}, 160401 (2002)
\bibitem{Gra} S.R. Granade, M.E. Gehm, K.M. O'Hara, J.E. Thomas,
{\it Phys. Rev. Lett.} {\bf 88}, 120405 (2002)
\bibitem{Roa} G. Roati, F. Riboli, G. Modugno, M. Inguscio,
{\it Phys. Rev. Lett.} {\bf 88}, 150403 (2002)
\bibitem{Schr} F. Schreck,
{\it Ann. Phys. Fr.} {\bf 28}, No. 2 (2003)
\bibitem{Tie} E. Tiesinga, B.J. Verhaar and H.T.C. Stoof,
{\it Phys. Rev. A} {\bf 47}, 4114 (1993)
\bibitem{Com} R. Combescot, {\it Phys. Rev. Lett.} {\bf 83}, 3766 (1999)
\bibitem{Holl} M. Holland, S.J.J.M.F. Kokkelmans, M.L. Chiofalo, R. Walser,
{\it Phys. Rev. Lett.} {\bf 87}, 120406 (2001)
\bibitem{Tim} E. Timmermans, K. Furuya, P.W. Milonni, A.K. Kerman,
{\it Phys. Lett. A} {\bf 285}, 228 (2001)
\bibitem{Oha} Y. Ohashi and A. Griffin,
{\it Phys. Rev. Lett.} {\bf 89}, 130402 (2002)
\bibitem{Bl1} M. Greiner, O. Mandel, T. Esslinger, T.W. H\"ansch, I. Bloch,
{\it Nature} {\bf 415}, 39-44 (2002)
\end{thebibliography}
\end{document}